\documentclass[11pt]{article}
\usepackage{amsmath,amssymb,amsthm,amsxtra,overpic,bbm,bm,epsfig,subfigure}
\usepackage{color}
\usepackage{slashed,stmaryrd}

\def\thefootnote{\fnsymbol{footnote}}

\begin{document}

\vspace{0.2cm}

\begin{center}
{\Large\bf Electroweak Vacuum Stability and Diphoton Excess at 750 GeV}
\end{center}

\vspace{0.2cm}

\begin{center}
{\bf Jue Zhang $^{a}$} \footnote{E-mail: zhangjue@ihep.ac.cn}
\quad {\bf Shun Zhou $^{a,~b}$} \footnote{E-mail: zhoush@ihep.ac.cn}
\\
{$^a$Institute of High Energy Physics, Chinese Academy of
Sciences, Beijing 100049, China \\
$^b$Center for High Energy Physics, Peking University, Beijing 100871, China}
\end{center}

\vspace{1.5cm}
\begin{abstract}
Recently, both ATLAS and CMS collaborations at the CERN Large Hadron Collider (LHC) have announced their observations of an excess of diphoton events around the invariant mass of $750~{\rm GeV}$ with a local significance of $3.6\sigma$ and $2.6\sigma$, respectively. In this paper, we interpret the diphoton excess as the on-shell production of a real singlet scalar in the $pp \to S \to \gamma \gamma$ channel. To accommodate the observed production rate, we further introduce a vector-like fermion $F$, which is carrying both color and electric charges. The viable regions of model parameters are explored for this simple extension of the Standard Model (SM). Moreover, we revisit the problem of electroweak vacuum stability in the same scenario, and find that the requirement for the electroweak vacuum stability up to high energy scales imposes serious constraints on the Yukawa coupling of the vector-like fermion and the quartic couplings of the SM Higgs boson and the new singlet scalar. Consequently, a successful explanation for the diphoton excess and the absolute stability of electroweak vacuum cannot be achieved simultaneously in this economical setup.
\end{abstract}

\begin{flushleft}
\hspace{0.8cm} PACS number(s): 12.38.Bx, 12.60.Fr, 14.80.Bn
\end{flushleft}

\def\thefootnote{\arabic{footnote}}
\setcounter{footnote}{0}

\newpage

\section{Motivation}

Exactly four years ago, the ATLAS~\cite{ATLAS:2012ae} and CMS~\cite{Chatrchyan:2012tx} collaborations announced publicly the experimental indications of a scalar particle $H$ in the narrow mass range of $124~{\rm GeV} < m^{}_H < 126~{\rm GeV}$ both in the diphoton channel $H \to \gamma\gamma$ and in the four-lepton mode $H \to Z Z^* \to 4 l$. Only half a year later, with higher integrated luminosities and more accumulated data, both collaborations~\cite{Aad:2012tfa,Chatrchyan:2012xdj} confirmed the existence of a scalar particle of mass $M^{}_H \approx 125~{\rm GeV}$, which is very likely to be the Higgs boson in the SM~\cite{Glashow:1961tr}. The Higgs boson is playing a particularly important role in the SM~\cite{Higgs:1964ia}, as it is responsible for the spontaneous gauge symmetry breaking of ${\rm SU(3)}^{}_{\rm C} \otimes {\rm SU(2)}^{}_{\rm L} \otimes {\rm U(1)}^{}_{\rm Y} \to {\rm SU(3)}^{}_{\rm C} \otimes {\rm U(1)}^{}_{\rm em}$ and for generating the masses of weak gauge bosons, quarks and charged leptons. However, the discovery of the SM-like Higgs boson is not the end of story, since there are several unsolved fundamental problems that are associated with such a scalar particle.

First of all, the radiative corrections to the Higgs-boson mass are not protected by any symmetries, so huge corrections from new physics at a high-energy scale, e.g., the scale $\Lambda^{}_{\rm GUT} \approx 10^{16}~{\rm GeV}$ of Grand Unified Theories (GUT) or the Planck scale $\Lambda^{}_{\rm Pl} \approx 10^{19}~{\rm GeV}$, should be cancelled out by an equally large counter term~\cite{'tHooft:1979bh}. There already exist possible solutions to this fine-tuning or naturalness problem, such as the Coleman-Weinberg mechanism~\cite{Coleman:1973jx}, which attempts to remove all the dimensionful parameters and restore a classically scale-invariant theory, or supersymmetries~\cite{Nilles:1983ge}, which guarantee an elegant cancellation between the radiative corrections from fermions and those from bosons.

Second, the scalar potential of the SM may develop another much deeper minimum at a large field value, destabilizing the electroweak vacuum~\cite{Cabibbo:1979ay,Altarelli:1994rb}. This indeed happens when the Higgs quartic coupling $\lambda^{\rm SM}$ becomes negative at a high-energy scale. In particular, for the Higgs-boson mass $m^{}_H \approx 125~{\rm GeV}$, the renormalization-group (RG) evolution of $\lambda^{\rm SM}$ is significantly affected by a large and negative contribution from the top-quark Yukawa coupling $y^{}_t \sim {\cal O}(1)$. In Ref.~\cite{Xing:2011aa}, it has been found that new physics should show up below a cutoff scale of $\Lambda^{}_{\rm VS} \approx 10^{12}~{\rm GeV}$ to stabilize the electroweak vacuum, for the Higgs-boson mass $m^{}_H = 125~{\rm GeV}$, the top-quark mass $M^{}_t = 172.9~{\rm GeV}$ and the strong fine-structure constant $\alpha^{}_{\rm s}(M^{}_Z) = 0.1184$ at the scale of the weak $Z$-boson mass $M^{}_Z = 91.2~{\rm GeV}$. The problem of vacuum stability has also been investigated by many other authors~\cite{Holthausen:2011aa}, and the improvements have been made by including high-order corrections~\cite{Degrassi:2012ry}. The final verdict is that the absolute stability of the SM vacuum up to the Planck scale is excluded at the $98\%$ confidence level for $m^{}_H < 126~{\rm GeV}$~\cite{Degrassi:2012ry}.

In addition, most free parameters in the SM arise from the Yukawa couplings between the Higgs boson and the fermions. The observed strong quark mass hierarchy, flavor mixing angles and CP violation remain unexplained and have been long-standing flavor puzzles~\cite{Xing:2014sja}. The solutions to those fundamental problems, together with the origin of neutrino masses~\cite{Zhou:2015kqs}, definitely call for new physics beyond the SM.

Very recently, based on the data collected from the LHC Run II at a center-of-mass energy $\sqrt{s} = 13~{\rm TeV}$, the ATLAS~\cite{ATLAS1512} and CMS~\cite{CMS:2015dxe} collaborations released their observations of a diphoton excess around the invariant mass of $750~{\rm GeV}$ with a local significance of $3.6~\sigma$ and $2.6~\sigma$, respectively. If these observations are confirmed by the future data, we can claim the first and direct discovery of new physics beyond the SM. Therefore, it is intriguing to investigate possible explanations for the diphoton excess, as done in a torrent of papers~\cite{Mambrini:2015wyu}. Different from most of the previous works, we focus on the problem of electroweak vacuum stability, and examine whether the new physics indicated by the diphoton exceess could help stabilize the electroweak vacuum. Although only a minimal extension of the SM with a real singlet scalar $S$ and a vector-like fermion $F$ is considered for illustration, the generalization of our analysis to other models proposed in Ref.~\cite{Mambrini:2015wyu} is straightforward.

The rest of our paper is structured as follows. In Sec. 2, the minimal model is introduced to explain the diphoton excess as a scalar resonance in $pp \to S \to \gamma\gamma$, where the vector-like fermion $F$ is implemented to mediate the $S \to gg$ and $S \to \gamma\gamma$ processes at the one-loop level. The diphoton signal rate and other collider constraints are used to explore the viable parameter space. In Sec. 3, we derive the conditions for the absolute vacuum stability, and present the RG equations (RGEs) for the SM gauge couplings, the Yukawa coupling of the vector-like fermion and quartic scalar couplings. Then, we have carried out a complete numerical study of model parameters and found that a successful explanation for the diphoton excess at $750~{\rm GeV}$ and the vacuum stability up to the Planck scale cannot be achieved simultaneously in this minimal model. Finally, we summarize in Sec. 4.

\section{The Diphoton Excess}

Motivated by the diphoton excess, we extend the SM with a real singlet scalar $S$ and a vector-like fermion $F$. Under the SM gauge group $S$ is singlet while $F$ transforms as $(R_{\rm C}^{},  R_{\rm W}^{})_{Y_{\rm F}^{}}$, where the color, weak isospin and hypercharge assignments are indicated in the conventional form, e.g., $R^{}_{\rm C}$ or $R^{}_{\rm W}$ stands for the dimension of the corresponding representation.  The general gauge-invariant Lagrangian can be written as
\begin{eqnarray}\label{eq:lagrangian}
{\cal L} \supset \frac{1}{2}(\partial^\mu S) (\partial^{}_\mu S) + {\rm i} \overline{F} \gamma^\mu D_\mu F - \left(y^{}_S \overline{F^{}_{\rm L}} F^{}_{\rm R} S + {\rm h.c.}\right) + V^{}_0(H, S) \; ,
\end{eqnarray}
where the covariant derivative is defined as $D^{}_\mu \equiv \partial^{}_\mu - {\rm i}g^{}_3 T^a(R^{}_{\rm C}) G^a_\mu - {\rm i}g^{}_2 t^i(R^{}_{\rm W}) W^i_\mu - {\rm i}g^\prime (Y^{}_{\rm F}/2) B^{}_\mu$ with $T^a(R^{}_{\rm C})$ and $t^i(R^{}_{\rm W})$ being proper representations of the generators of ${\rm SU(3)}^{}_{\rm C}$ and ${\rm SU(2)}^{}_{\rm L}$ gauge groups, respectively. The electric charge of $F$ is then given by the Gell-Mann-Nishijima formula $Q^{}_{\rm F} = I^3_{\rm W} + Y^{}_{\rm F}/2$.

In addition, the tree-level scalar potential can be parametrized as follows
\begin{eqnarray}\label{eq:V0}
V^{}_0(H, S) = \lambda^{}_H \left(H^\dagger H - \frac{v^2}{2}\right)^2 + \frac{\lambda^{}_S}{4} \left(S^2 - w^2\right)^2 + \lambda^{}_{HS} \left(H^\dagger H - \frac{v^2}{2}\right) \left(S^2 - w^2\right) \; ,
\end{eqnarray}
where the vacuum expectation values (vev's) of the SM Higgs and new singlet scalar fields are $\langle H \rangle = v/\sqrt{2}$ and $\langle S \rangle = w$. Note that we have imposed a $Z^{}_2$ symmetry on the Lagrangian in Eq.~(\ref{eq:lagrangian}), under which all the SM fields are even, while the other fields transform as $F^{}_{\rm L} \to F^{}_{\rm L}$ and $F^{}_{\rm R} \to -F^{}_{\rm R}$ and $S \to -S$. After the spontaneous symmetry breaking, the vector-like fermion acquires its mass $m^{}_F = y^{}_S w$. Some comments on the model are in order. The spontaneous breaking of the $Z^{}_2$ symmetry may lead to the domain-wall problem, which however can be solved by adding an explicit symmetry-breaking term into the Lagrangian. For instance, a Dirac mass term $\tilde{m}^{}_F \overline{F^{}_{\rm L}} F^{}_{\rm R}$ suffices for this purpose. In this case, the fermion mass is given by $m^{}_F = \tilde{m}^{}_F + y^{}_S w$, so we may just take both $m^{}_F$ and $y^{}_S$ as free model parameters. This is also true for a more general theory in which the fermion mass is not subject to spontaneous symmetry breaking but some underlying strong dynamics. In the present work, we take the scalar potential in Eq.~(\ref{eq:V0}) as a phenomenological example, which could be realized in various theoretical models with a Higgs portal to new physics~\cite{DiChiara:2014wha,McDonald:1993ex}.

\subsection{Signals and Constraints}

After specifying the theoretical framework, we are now in a good position to explain the diphoton excess at $750~{\rm GeV}$ and examine the existing constraints on the minimal model presented in Eqs.~(\ref{eq:lagrangian}) and (\ref{eq:V0}). The signal cross section can be estimated as $\sigma(pp \to \gamma \gamma) = (10\pm 3)~{\rm fb}$ for ATLAS and $(6\pm 3)~{\rm fb}$ for CMS at the center-of-mass energy of $\sqrt{s} = 13~{\rm TeV}$, which are compatible at the $2\sigma$ level with the non-observation of diphoton excess in the LHC Run I data at $\sqrt{s} = 8~{\rm TeV}$~\cite{8TeV_gamma_gamma_ATLAS,8TeV_gamma_gamma_CMS}. Apart from the hint at an invariant mass of $M \approx 750~{\rm GeV}$, the ATLAS collaboration also suggests a best-fit width $\Gamma$ of about $45~{\rm GeV}$ or equivalently $\Gamma/M \approx 0.06$, while the data from CMS favor mostly a narrow width. But a wide resonance is also consistent with the CMS data at the $2\sigma$ level~\cite{Franceschini:2015kwy}.

Now it becomes evident that the diphoton excess can be interpreted as the $s$-channel production of the singlet scalar $S$ of mass $m^{}_S = 750~{\rm GeV}$. Naively, we expect that the production is dominated by the gluon-gluon fusion $gg \to S$ and the decay is via $S \to \gamma \gamma$.
As demonstrated in Ref.~\cite{Knapen:2015dap}, just one singlet scalar $S$ coupled to the SM particles cannot explain this diphoton excess without violating existing constraints from other channels. It is therefore necessary to add the vector-like fermion $F$, which meditates both production and decay processes through one-loop diagrams. In order to avoid tight constraints from the diboson searches at the LHC, we take $F$ to be an ${\rm SU(2)}_{\rm L}^{}$ singlet, and its sole contributions to the partial decay widths of $S \rightarrow gg$ and $\gamma\gamma$ are found as follows \cite{Gunion:1989we}
\begin{eqnarray}\label{eq:decaywidths}
\Gamma_{S \rightarrow gg}^{F} &=& \frac{\alpha_s^2 m_S^3}{128\pi^3} \left | \frac{y_S^{}}{m_F^{}} A_{1/2}^{}(\tau_S^{}) \right |^2, \label{eq:decaygg} \\
\Gamma_{S \rightarrow \gamma\gamma}^{F} &=& \frac{\alpha_e^2 m_S^3}{1024\pi^3} \left | 2 R_{\rm C}^{} Q_F^2 \frac{y_S^{}}{m_F^{}} A_{1/2}^{}(\tau_S^{}) \right |^2,
\label{eq:decayaa}
\end{eqnarray}
where the fine structure constants $\alpha^{}_{\rm s}(\mu) \equiv g^2_3(\mu)/(4\pi)$ and $\alpha(\mu) \equiv e^2(\mu)/(4\pi)$ are both evaluated at $\mu = 750~\mathrm{GeV}$. In addition, we have defined $\tau_S^{} = 4 m_F^2/m_S^2$, and the function $A_{1/2}(\tau)$ is
\begin{eqnarray} \label{eq:Ahalf}
A_{1/2}^{}(\tau) = -2\tau \left[1 + (1-\tau) f(\tau) \right],
\end{eqnarray}
with $f(\tau)$ given by
\begin{eqnarray}
f(\tau) &=& \begin{cases}
\left( \sin^{-1}\sqrt{\displaystyle \frac{1}{\tau}} \right)^2, & \tau \geq 1 \\
~ & ~
\\
-\displaystyle \frac{1}{4} \left( \displaystyle \ln \frac{1+\sqrt{1-\tau}}{1-\sqrt{1-\tau}} - \mathrm{i} \pi \right)^2, & \tau < 1
\end{cases}.
\end{eqnarray}
The partial decay widths to $ZZ$ and $Z\gamma$ via the vector-like fermions are found to correlate with that to $\gamma\gamma$, however, they are respectively suppressed by $2 \tan^2\theta_W^{}$ and $\tan^4\theta_W^{}$, where $\theta_W^{}$ is the Weinberg angle.

Through the interactions with the SM Higgs field, this scalar field $S$ acquires additional decay modes. First, according to the scalar potential in Eq.~(\ref{eq:V0}), a mixing between $S$ and the SM Higgs boson occurs after the spontaneous symmetry breaking. Because of the mixing with $H$, the singlet scalar $S$ is coupled to the SM fermions and gauge bosons, although the coupling strengths are suppressed by the mixing angle $\theta$.\footnote{As we will show in next section, $\theta$ can be expressed in terms of the model parameters. For the moment, we keep it as general as possible and investigate its viable values.} Since a SM-like Higgs boson with a mass of $750~\mathrm{GeV}$ predominantly decays into $WW$, $ZZ$ and $t\bar{t}$ \cite{Heinemeyer:2013tqa}, we thus expect contributions from the mixing with the SM Higgs field to the partial decay widths of $S \rightarrow WW, ZZ$ and $t\bar{t}$. Their contributions can be easily found by multiplying the corresponding partial decay widths of a SM-like Higgs boson of mass $750~\mathrm{GeV}$ with a suppression factor of $\sin^2\theta$.

Apart from the above mixing effect, the scalar potential in Eq.~(\ref{eq:V0}) also induces the decay of $S$ into two SM Higgs bosons through the quartic coupling $\lambda_{HS}^{} S^2 H^\dagger H$, when $S$ picks up a nonzero vev $w$. A precise study of this decay mode involves changing the basis to the mass eigenstates of both $S$ and $H$, i.e., $s$ and $h$, and carefully taking into account the singlet-doublet mixing effect. See, e.g., Refs.~\cite{Bowen:2007ia,Bojarski:2015kra}, for more details. However, given the fact that the mixing angle $\theta$ tends to be small in order to satisfy the experimental constraints, we can write the trilinear interaction as $a h^2 s$ at leading order, where the coupling $a$ is approximately given by $a \approx \lambda_{HS}^{} w$. From the later discussion on the diagalization of the scalar mass matrix, we may further write $a$ in terms of the mass of the scalar field $m_S^{}$ and the mixing angle $\theta$ when $v \ll w$, namely, $a \approx m_S^2 \sin\theta /v$. Then, according to Ref.~\cite{Bowen:2007ia}, we find the partial decay width of $S \rightarrow hh$ as follows
\begin{eqnarray}
\Gamma_{S \rightarrow hh}^{} \approx \frac{\sin^2\theta m_S^3}{8\pi v^2} \sqrt{1-\frac{4 m_H^2}{m_S^2}},
\end{eqnarray}
where $m_H^{} \simeq 125~{\mathrm{GeV}}$ is the mass of the SM Higgs boson.

To summarize, due to the interactions with the vector-like fermion $F$ and the SM Higgs boson $H$, the scalar boson $S$ decays in several modes, e.g., $S \rightarrow gg, \gamma\gamma, ZZ, Z\gamma, WW, t\bar{t}$ and $hh$.
It should be noted that among these decay modes, $S \rightarrow gg, \gamma\gamma, ZZ$ and $Z\gamma$ contain interference effects between the contributions from the vector-like fermion and that from the SM, which have been taken into account in our calculations. Moreover, we also include the next-to-leading order $K$-factor of $1+67\alpha_s^{}/(4\pi)$ for the $S \rightarrow gg$ decay mode.

Given the partial decay widths of $S$, we are now ready to calculate its diphoton production rate at the LHC. For the resonant production, the signal cross section can be expressed in terms of relevant decay widths corresponding to the production and decay processes. At $\sqrt{s} = 13~{\rm TeV}$, the diphoton signal rate can be recast into the form~\cite{Franceschini:2015kwy}
\begin{eqnarray} \label{eq:diphoton_rate}
\frac{\Gamma_{\rm prod}^{} \Gamma_{S \rightarrow \gamma\gamma}^{}}{\Gamma_{\rm tot}^{} M_S^{}} \simeq 1.1 \times 10^{-6},
\end{eqnarray}
where $\Gamma^{}_{\rm tot}$ is the total decay width of $S$, and $\Gamma_{\rm prod}$ is the decay width corresponding to the dominant production mechanism of $S$, namely, $\Gamma^{}_{\rm prod} = \Gamma^{}_{S \to gg}$.

In addition to the condition in Eq.~(\ref{eq:diphoton_rate}) that accounts for the correct diphoton production rate, we should also ensure that the production rates of final states other than two photons are not in contradiction with current experimental observations. To this end, we quote the results from Ref.~\cite{Franceschini:2015kwy}, in which the present constraints from direct searches for $pp \to XX$ have been reinterpreted as ratios of partial decay widths $\Gamma^{}_{S \to XX}$ to $\Gamma_{S \rightarrow \gamma\gamma}^{}$. Those constraints are shown in Table \ref{tb:constraints}, where the production rates at $\sqrt{s} = 13~{\rm TeV}$ are scaled from those at $\sqrt{s} = 8~{\rm TeV}$ by a factor of $r = \sigma_{13\mathrm{TeV}}/\sigma_{8\mathrm{TeV}} \approx 5$, and the decay width $\Gamma^{}_{S \to \gamma \gamma}$ fits to the central value of the observed diphoton excess.
\begin{table}
\centering
\begin{tabular}{cc}
\hline
\hline
Final States $XX$  & Bounds on $\Gamma_{S \rightarrow XX}^{} / \Gamma_{S \rightarrow \gamma\gamma}^{}$ \cite{Franceschini:2015kwy} \\
\hline
$WW$ & $< 20 (r/5)$ \\
$ZZ$ & $< 6 (r/5) $ \\
$Z\gamma$ & $<2 (r/5) $\\
$t\bar{t}$ & $< 300 (r/5) $ \\
$j j$ & $ < 1300 (r/5) $ \\
$h h$ & $ < 20 (r/5) $ \\
\hline
\hline
\end{tabular}
\caption{Upper bounds on partial decay widths $\Gamma^{}_{S \to XX}$ to various final states normalized by $\Gamma_{S \rightarrow \gamma\gamma}^{}$, assuming that the production cross section grows as $r = \sigma_{13\mathrm{TeV}}/\sigma_{8\mathrm{TeV}} \approx 5$, and that $S \rightarrow \gamma\gamma$ fits the central value of the observed diphoton excess. }
\label{tb:constraints}
\end{table}

\subsection{Viable Parameter Space}

Next, using the signal rate in Eq.~(\ref{eq:diphoton_rate}) and the constraints in Table~\ref{tb:constraints}, we explore the allowed parameter space of our model. The details of our numerical calculations can be summarized as follows:
\begin{itemize}
\item For simplicity, we assume that the vector-like fermion forms a color triplet, belonging to the fundamental representation of the ${\rm SU(3)}^{}_{\rm C}$ gauge group, i.e., $R^{}_{\rm C} = {\bf 3}$. The generalization to a different color multiplet is straightforward.

\item We focus on the parameter space of the mixing angle $\sin \theta$ and the Yukawa coupling $y^{}_S$, which are of crucial importance for our later discussions on the electroweak vacuum stability in the next section. For the values of $\sin \theta$, we choose a constrained region of $10^{-3} < \sin \theta < 0.32$, where the upper bound arises from current measurements of Higgs couplings~ \cite{Falkowski:2015iwa}. As for the Yukawa coupling $y^{}_S$, the range is chosen as $10^{-2} < y^{}_S < \sqrt{4\pi}$, for which the upper bound is required by the perturbativity.

\item Finally, for a given set of $\sin\theta$ and $y^{}_S$, we solve Eq.~(\ref{eq:diphoton_rate}) for the fermion mass $m^{}_F$, and  further restrict it into the range of $[0.4, 10]~{\rm TeV}$. Here we consider a lower bound of $m^{}_F = 400~{\rm GeV}$ so as to employ the threshold effect presented in $A_{1/2}^{}(\tau)$ to enhance the signal production rate. Vector-like fermions with such a light mass are actually already excluded by 8 TeV LHC results \cite{VLQ_limit}, if assuming that they can decay to the SM top or bottom quark via $W$, $Z$ or $H$. However, in our scenario the vector-like fermion $F$ has no such decay modes, so the bounds from current experimental searches are evaded. For the upper bound of $m^{}_F$ we choose it for the testability in future collider experiments.

\end{itemize}

\begin{figure}[!t]
\centering
\includegraphics[scale=0.55]{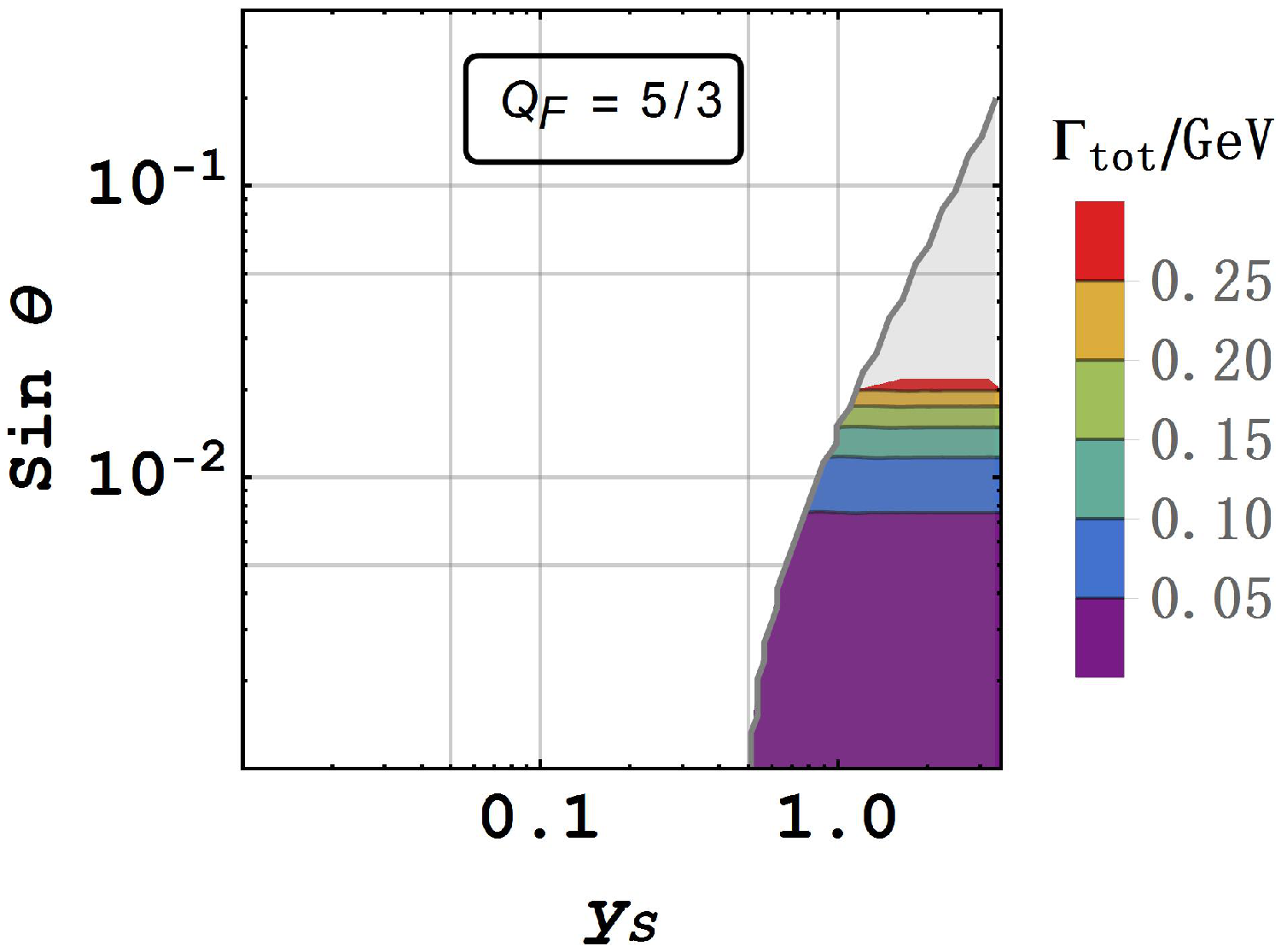}\quad
\includegraphics[scale=0.55]{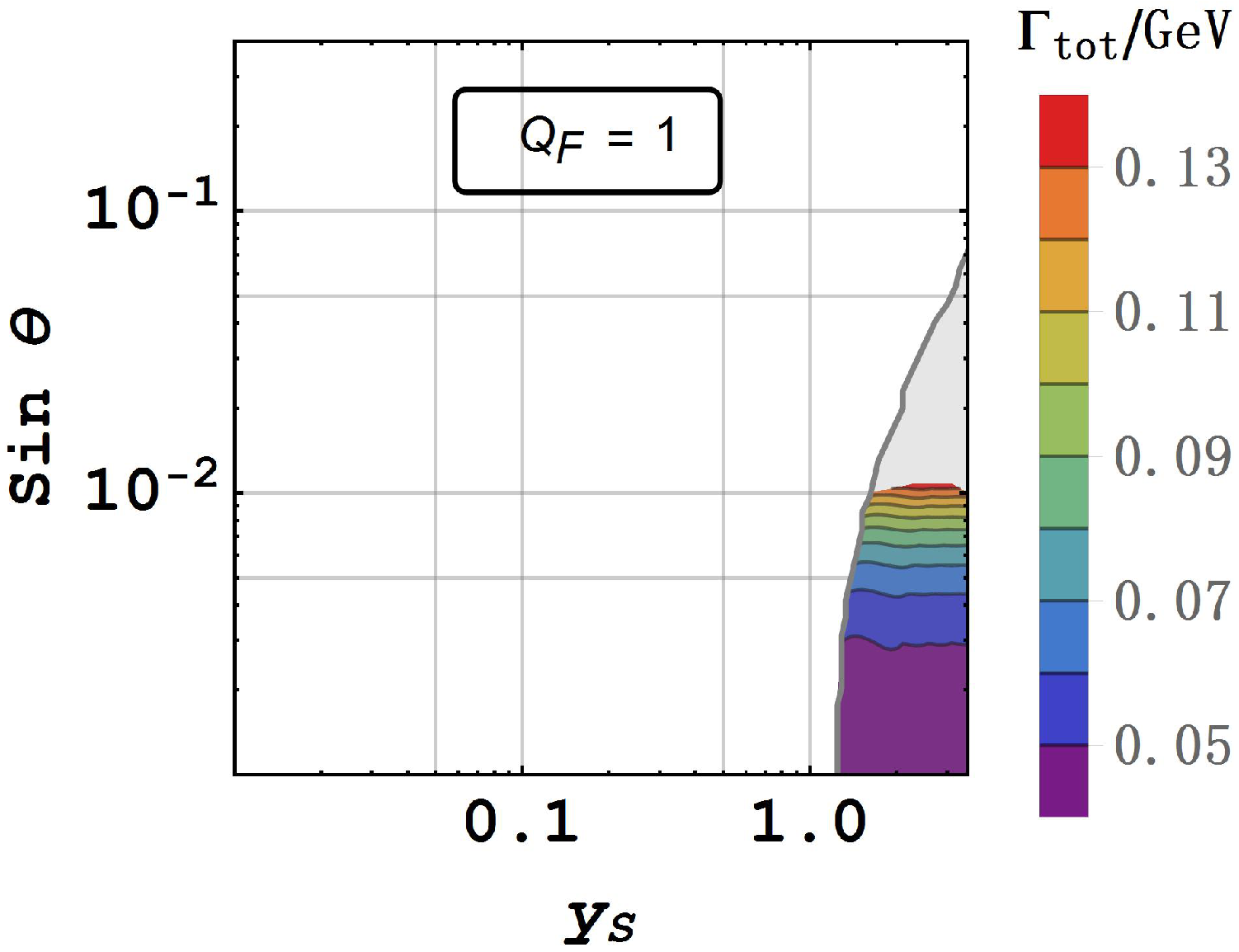}
\caption{Allowed parameter space for $\sin\theta$ and $y_S^{}$ from diphoton excess and other collider constraints. Two different electric charges of $Q_F^{} = 5/3$ (left) and $Q_F^{} = 1$ (right) are assumed. The gray shaded regions indicate the allowed parameter space when only the diphoton production rate condition of Eq.~(\ref{eq:diphoton_rate}) is considered, while further imposing the constraints from other decay modes leads to the colored regions. Labeled contours are to indicate the total decay width $\Gamma_{\mathrm{tot}}$ of the scalar $S$.}
\label{fg:LHC_constraints}
\end{figure}

Following the above strategy, we obtain the allowed parameter space for $\sin\theta$ and $y_S^{}$ in Fig.~\ref{fg:LHC_constraints}, where two possible electric charge assignments of $Q_F^{} = 5/3$ and $1$ are assumed for illustration.\footnote{If the electric charge of new vector-like fermions is much larger than $Q^{}_F = 5/3$, the electromagnetic gauge coupling will run quickly into a non-perturbative region at the TeV scale. The requirement for perturbativity has been taken into account in the next section, and also discussed by others~\cite{g1_running}.} The gray shaded regions indicate the allowed parameter space when only the diphoton production rate condition of Eq.~(\ref{eq:diphoton_rate}) is considered, while further imposing the constraints from other decay modes leads to the colored regions. Notice that vector-like fermions with exotic electric charges may appear in the composite models~\cite{Franceschini:2015kwy}, and the studies of the scenarios with different electric charges can be performed in a similar way. Two comments on the numerical results are in order:
\begin{enumerate}
\item As one can observed from Fig.~\ref{fg:LHC_constraints}, to account for the observed diphoton rate, we need a rather large value of $y_S^{}$ and a small mixing angle $\theta$. This observation is understandable since a large signal rate needs a big Yukawa coupling and a sizable electric charge. Therefore, if a smaller electric charge $Q^{}_F$ is taken, one has to increase the Yukawa coupling $Y^{}_S$ or multiply the generations of vector-like fermions in order to explain the diphoton excess.

\item The color bars in Fig.~\ref{fg:LHC_constraints} indicate the total decay width of $S$ is in general narrow. In both cases the widths are only on the order of sub-GeV, which is much narrower than the observed $\Gamma = 45~{\rm GeV}$. However, for the time being, this is not a big issue, since a narrow width cannot be excluded. On the theoretical side, this is reasonable for a weakly-coupled theory, whereas a wide resonance is naturally expected in a strongly-coupled theory, such as the composite models.
\end{enumerate}
Finally, it is worthwhile to note that the fermion mass $m^{}_F$ and the Yukawa coupling $y^{}_S$ can be closely related, as we mentioned before. In this case, the model parameters will be more severely constrained.

\section{Electroweak Vacuum Stability}

If the diphoton excess is further confirmed by future data, one will be convinced that new physics should be introduced to account for the observations, as we have shown. If the minimal scenario of a singlet scalar and a vector-like fermion works well for this purpose, then an interesting question is whether they help stabilize the electroweak vacuum. The same question can also be asked in any other new physics scenarios. Unfortunately, as we demonstrate below, it is impossible to simultaneously explain the diphoton excess and solve the vacuum stability problem in this minimal scenario. Even worse, the new fermion $F$ with a sizable Yukawa coupling $y^{}_S$ will destabilize the electroweak vacuum.

\subsection{Conditions for Absolute Stability}

Since the SM Higgs boson of mass $m^{}_H \approx 125~{\rm GeV}$ is relatively light, while the top quark is heavy, the quartic Higgs coupling $\lambda^{\rm SM}$ may run into a region of negative values at a high renormalization scale. As a consequence, the effective potential becomes unbounded from below, or a global minimum that is much deeper than the electroweak vacuum develops at large field values. For $m^{}_H = 125~{\rm GeV}$, $M^{}_t = 172.9~{\rm GeV}$ and $\alpha^{}_{\rm s}(M^{}_Z) = 0.1184$, it has been found that a negative value of $\lambda^{}_H(\Lambda^{}_{\rm VS})$ is reached for $\Lambda^{}_{\rm VS} \approx 4\times 10^{12}~{\rm GeV}$~\cite{Xing:2011aa}.

In Ref.~\cite{EliasMiro:2012ay}, a simple but intriguing solution to the instability problem of the electroweak vacuum has been proposed. The essential idea is to introduce an extremely heavy scalar of mass $m^{}_S = 10^9~{\rm GeV}$, which leads to a threshold effect on the quartic coupling of Higgs boson. From the potential in Eq.~(\ref{eq:V0}), one can immediately derive the mass matrix for the scalars
\begin{eqnarray}\label{eq:massmatrix}
{\cal M}^2 = 2 \left(\begin{matrix} \lambda^{}_H v^2 & \lambda^{}_{HS} v w \cr \lambda^{}_{HS} v w & \lambda^{}_S w^2 \end{matrix}\right) \; ,
\end{eqnarray}
from which one can further obtain the scalar masses
\begin{eqnarray}\label{eq:matching_mass}
m^2_H &=& \lambda^{}_H v^2 + \lambda^{}_S w^2 - \sqrt{\left(\lambda^{}_S w^2 - \lambda^{}_H v^2\right)^2 + 4\lambda^2_{HS} v^2 w^2} \; , \nonumber \\
m^2_S &=& \lambda^{}_H v^2 + \lambda^{}_S w^2 + \sqrt{\left(\lambda^{}_S w^2 - \lambda^{}_H v^2\right)^2 + 4\lambda^2_{HS} v^2 w^2} \; .
\end{eqnarray}
In Ref.~\cite{EliasMiro:2012ay}, both $m^{}_S$ and $w$ are assumed to be much larger than the electroweak scale, so the mixing between singlet and doublet scalar bosons is extremely small (i.e., $v/w \ll 1$) and the scalar masses turn out to be $m^2_H = 2v^2 (\lambda^{}_H - \lambda^2_{HS}/\lambda^{}_S)$ and $m^2_S = 2\lambda^{}_S w^2 + 2(\lambda^2_{HS}/\lambda^{}_S) v^2$. Therefore, even for $w \to \infty$, there exists a constant shift in the quartic Higgs coupling $\lambda^{}_H \to \lambda^{}_H - \lambda^2_{HS}/\lambda^{}_S$. It is this threshold effect that has been implemented in Ref.~\cite{EliasMiro:2012ay} to eliminate the instability problem.

Different from the authors of Ref.~\cite{EliasMiro:2012ay}, we fix the scalar masses at those observed from the ATLAS and CMS experiments, namely, $m^{}_H \approx 125~{\rm GeV}$ and $m^{}_S \approx 750~{\rm GeV}$. In this case, the mixing between doublet and singlet scalars can be significant
\begin{eqnarray}\label{eq:matching_angle}
\tan 2\theta = \frac{2 \lambda^{}_{HS} v w}{\lambda^{}_S w^2 - \lambda^{}_H v^2} \; ,
\end{eqnarray}
depending on the vev's and quartic scalar couplings. As shown in the previous section, the mixing angle $\theta$ will be severely constrained from collider experiments.

In order to find out the conditions for the absolute stability, we recast the scalar potential into the following form
\begin{eqnarray}\label{eq:condition}
V^{}_0(H,S) = \lambda^{}_H \left[\left(H^\dagger H - \frac{v^2}{2}\right) + \frac{\lambda^{}_{HS}}{2\lambda^{}_H} \left(S^2 - w^2\right)\right]^2 +\frac{1}{4} \left(\lambda^{}_S - \frac{\lambda^2_{HS}}{\lambda^{}_H}\right) \left(S^2 - w^2\right)^2 \; ,
\end{eqnarray}
from which one can see that the condition for $V^{}_0(H, S) > 0$ requires $\lambda^{}_H > 0$, $\lambda^{}_S > 0$ and $\lambda^{}_S \lambda^{}_H > \lambda^2_{HS}$. In the following discussions, we implement the latter three inequalities as the criteria for an absolute electroweak vacuum. Some remarks are helpful. First, the parametrization of $V^{}_0(H,S)$ ensures that the SM vacuum corresponds to $V^{}_0 = 0$, so the requirement of $V^{}_0(H, S) > 0$ at large values of scalar fields guarantees that there is no minimum deeper than the SM vacuum. Second, to reliably compute the scalar potential at a large field value, one has to evaluate the quartic couplings at the corresponding energy scale. Therefore, the RGEs for the gauge, Yukawa and quartic couplings should be derived. This is the main task for the next subsection.

\subsection{Renormalization-Group Equations}

In this subsection, we derive the one-loop RGEs for the dimensionless couplings in the extension of the SM with a singlet real or complex scalar and a general vector-like fermion, whose quantum numbers under the SM gauge group $(R^{}_{\rm C}, R^{}_{\rm W})^{}_{Y^{}_{\rm F}}$ can be arbitrary. In Refs.~\cite{Kadastik:2009cu,Lerner:2009xg,Gonderinger:2009jp,Gonderinger:2012rd}, the one-loop RGEs for an additional scalar have been obtained, while no additional vector-like fermions are considered. A similar scenario to ours has been studied in Ref.~\cite{Xiao:2014kba}, where the authors investigated a specific vector-like fermion model and provided relevant RGEs for the real scalar case. Here we extend the RGE study in Ref.~\cite{Xiao:2014kba} in two aspects. First, both real and complex scalars are examined. Second, the most general assignment of SM gauge quantum numbers to the vector-like fermions is studied, so that other vector-like fermion models can be covered as well. In practice, we adopt the general one-loop RGE formulas given in Ref.~\cite{RGE_general}.

Let us start with the real scalar case. First of all, the RGEs of three gauge couplings get modified because of the additional vector-like fermions that also are carrying SM gauge quantum numbers. Above the mass threshold of these vector-like fermions, we find\footnote{A GUT normalization of $g_1^{} = \sqrt{5/3} g^\prime$ is used.}
\begin{eqnarray}
16\pi^2_{} \frac{{\rm d}g_i^{}}{{\rm d} \ln \mu} = - b_i^{} g_i^3, \quad (i = 1,2,3)
\end{eqnarray}
with the coefficients of $b_i^{}$ given by
\begin{eqnarray}
b_1^{} &=& b_1^{\rm SM} - \frac{1}{5} R_{\rm C}^{} R_{\rm W}^{} Y_{\rm F}^2, \nonumber \\
b_2^{} &=& b_2^{\rm SM} - \frac{4}{3} R_{\rm C}^{} D_2^{}(R_{\rm W}^{}), \nonumber \\
b_3^{} &=& b_3^{\rm SM} - \frac{4}{3} R_{\rm W}^{} D_3^{}(R_{\rm C}^{}), \nonumber
\end{eqnarray}
where $ (b_1^{\rm SM}, b_2^{\rm SM}, b_3^{\rm SM}) = \left(-41/10, 19/6, 7 \right)$ are coefficients in the SM, and $D_n^{}(R)$ denotes the quadratic Dynkin index of the representation $R$ of the group ${\rm SU}(n)$. Here we take the convention that $D_n^{}({\bf n}^2-1) = n$ for the adjoint representation, while $D_n^{}(\mathbf{n})= D_n^{}(\overline{\mathbf{n}}) = 1/2$ for the fundamental representation, where $\mathbf{\overline{n}}$ is the conjugate representation of $\mathbf{n}$.

Next, we turn to the Yukawa couplings. Since the SM Higgs field does not couple to the extra vector-like fermions, as well as no additional Yukawa couplings between SM fermions and the extra scalar field, the RGEs of the SM Yukawa couplings stay the same. For the newly introduced Yukawa coupling $y_S^{}$, we obtain its RGE as
\begin{eqnarray}\label{eq:RGEcom0}
16\pi^2_{} \frac{{\rm d} y_S^{}}{{\rm d} \ln \mu} = y_S^{} \left\{ \left( 3 + 2 R_{\rm C}^{} R_{\rm W}^{} \right) y_S^2 - \left[ 6 g_3^2 C_3^{} (R_{\rm C}^{}) + 6 g_2^2 C_2^{} (R_{\rm W}^{}) + \frac{9}{10} Y_{\rm F}^2 g_1^2 \right]  \right\},
\end{eqnarray}
where $C_n^{}(R)$ is the quadratic Casimir invariants for the representation $R$ of the group ${\rm SU}(n)$. For example, we have $C_3^{}(\mathbf{3}) = C_3^{}(\mathbf{\overline{3}}) = 4/3$ and $C_2^{}(\mathbf{2})  = 3/4$.

Finally, we give the RGEs of the quartic couplings $\lambda_H^{}$, $\lambda_{HS}^{}$ and $\lambda_S^{}$. Neglecting all the SM Yukawa couplings except for that of the top quark $y_t^{}$, we find
\begin{eqnarray}\label{eq:RGEcom}
16\pi^2_{} \frac{{\rm d} \lambda_H^{}}{{\rm d} \ln \mu} &=& \left( 12 y_t^2 - \frac{9}{5} g_1^2 - 9 g_2^2\right) \lambda_H^{} - 6 y_t^4 + \frac{3}{8}\left[ 2 g_2^4 + \left( \frac{3}{5} g_1^2 + g_2^2 \right)^2 \right] + 24 \lambda_H^2 + 2 \lambda_{HS}^2 , \nonumber \\
16\pi^2_{} \frac{{\rm d} \lambda_{HS}^{}}{{\rm d} \ln \mu} &=& \frac{1}{2} \left( 12y_t^2 - \frac{9}{5} g_1^2 - 9 g_2^2 \right) \lambda_{HS}^{} + 6 \lambda_{HS}^{} ( 2\lambda_H^{} + \lambda_S^{}) + 8 \lambda_{HS}^2 + 4 R_{\rm C}^{} R_{\rm W}^{} y_S^2 \lambda_{HS}^{}, \nonumber \\
16\pi^2_{} \frac{{\rm d} \lambda_S^{}}{{\rm d} \ln \mu} &=& 8 \lambda_{HS}^2 + 18 \lambda_S^2 + 8 R_{\rm C}^{} R_{\rm W}^{} y_S^2 (\lambda_S^{} - y_S^2).
\end{eqnarray}
As one can see, the RGEs are coupled together, and the running of $\lambda_H^{}$ can now be remarkably affected by $\lambda_{HS}^{}$. Moreover, the additional Yukawa coupling $y_S^{}$ appears only in the RGEs of $\lambda_{HS}^{}$ and $\lambda_S^{}$, and it can play an important role in the vacuum stability study, as $\lambda_S^{}$ could be driven to be negative at some high energy scale if $y_S^{}$ were too large.

The complex scalar case can be studied in a similar way, and the corresponding RGEs differ from those in Eqs.~(\ref{eq:RGEcom0}) and (\ref{eq:RGEcom}) only in $y_S^{}$ and three quartic couplings. The Lagrangian takes the same form as in Eq.~(\ref{eq:lagrangian}) but with a distinct scalar potential
\begin{eqnarray}\label{eq:lag_complex}
V^{}_0(H, S) = \lambda^{}_H \left(H^\dagger H - \frac{v^2}{2}\right)^2 + \lambda^{}_S \left(S^\dagger S - \frac{w^2}{2}\right)^2 + 2\lambda^{}_{HS} \left(H^\dagger H - \frac{v^2}{2}\right) \left(S^\dagger S - \frac{w^2}{2}\right) \; .
\end{eqnarray}
In this case, we find the RGE of $y_S^{}$ as follows
\begin{eqnarray}\label{eq:complex_ys}
16\pi^2_{} \frac{{\rm d} y_S^{}}{{\rm d} \ln \mu} = y_S^{} \left\{ \left( 1 + R_{\rm C}^{} R_{\rm W}^{} \right) y_S^2 - \left[ 6 g_3^2 C_3^{} (R_{\rm C}^{}) + 6 g_2^2 C_2^{} (R_{\rm W}^{}) + \frac{9}{10} Y_{\rm F}^2 g_1^2 \right]  \right\},
\end{eqnarray}
while the RGEs of quartic couplings are given by
\begin{eqnarray}
16\pi^2_{} \frac{{\rm d} \lambda_H^{}}{{\rm d} \ln \mu} &=& \left( 12 y_t^2 - \frac{9}{5} g_1^2 - 9 g_2^2\right) \lambda_H^{} - 6 y_t^4 + \frac{3}{8}\left[ 2 g_2^4 + \left( \frac{3}{5} g_1^2 + g_2^2 \right)^2 \right] + 24 \lambda_H^2 + 4 \lambda_{HS}^2 , \nonumber \\
16\pi^2_{} \frac{{\rm d} \lambda_{HS}^{}}{{\rm d} \ln \mu} &=& \frac{1}{2} \left( 12y_t^2 - \frac{9}{5} g_1^2 - 9 g_2^2 \right) \lambda_{HS}^{} + 4 \lambda_{HS}^{} ( 3\lambda_H^{} + 2\lambda_S^{}) + 8 \lambda_{HS}^2 + 2 R_{\rm C}^{} R_{\rm W}^{} y_S^2 \lambda_{HS}^{}, \nonumber \\
16\pi^2_{} \frac{{\rm d} \lambda_S^{}}{{\rm d} \ln \mu} &=& 8 \lambda_{HS}^2 + 20 \lambda_S^2 + 2 R_{\rm C}^{} R_{\rm W}^{} y_S^2 (2\lambda_S^{} - y_S^2).
\end{eqnarray}
The applications of the RGEs in the case of complex scalar can be made similarly as we shall do in the next subsection for the real case. This indeed makes sense, as recently one also considers the explanation for the diphoton excess with a complex scalar, e.g., see Refs.~\cite{Cao:2016cok}.

\subsection{Viable Parameter Space}
\begin{figure}[!t]
\centering
\includegraphics[scale=1.0]{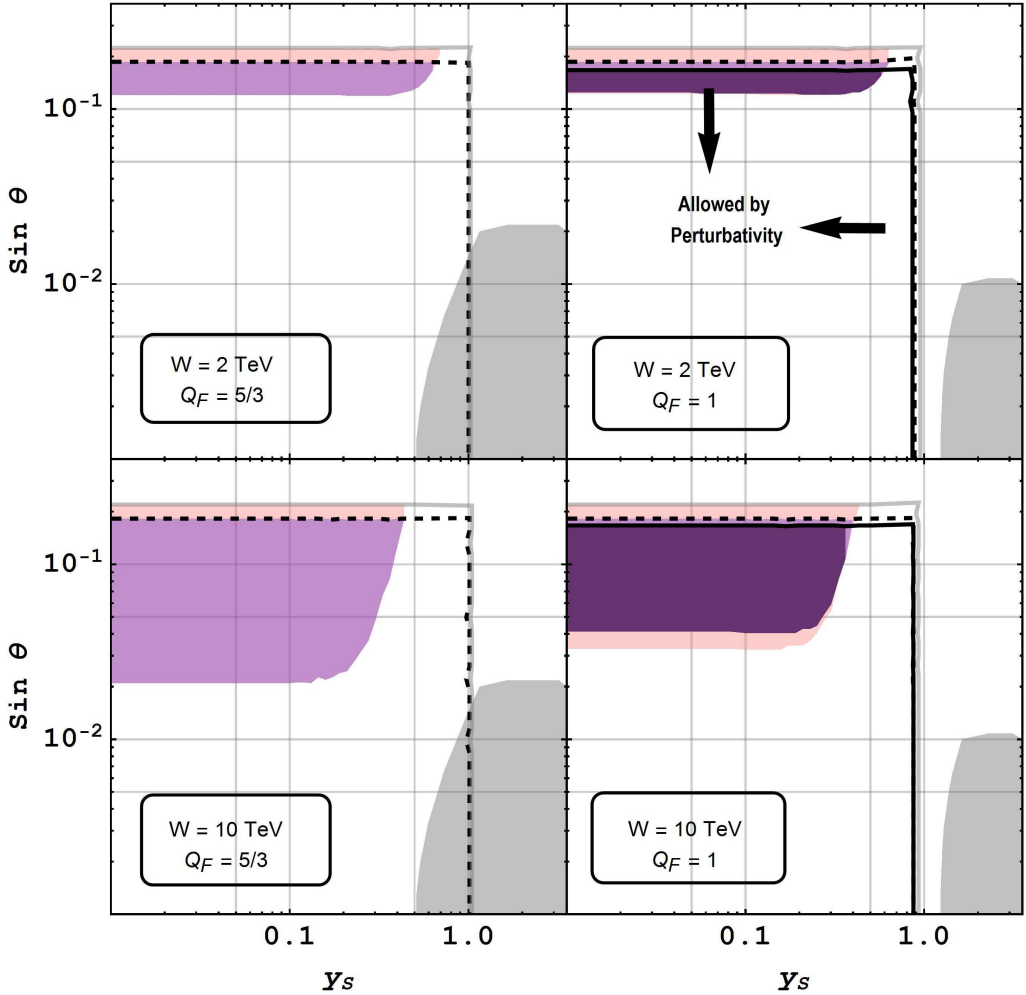}
\caption{Allowed parameter space for $y_S^{}$ and $\sin\theta$, where the electroweak vacuum stability is preserved till the Planck scale ($10^{19}~\mathrm{GeV}$, dark purple), the GUT scale ($10^{16}~\mathrm{GeV}$, dark purple + purple) or the same instability scale as in the SM ($10^{12}~\mathrm{GeV}$, dark purple + purple + pink).  Two different values $w = 2$ and $10~\mathrm{TeV}$ are taken for illustration, and $Q_F^{} = 5/3$ and $1$ are assumed as in Fig.~\ref{fg:LHC_constraints}. The boundaries of regions that satisfy the perturbativity requirements up to the Planck scale, the GUT scale and the same instability scale as in the SM are indicated by the black, dashed black and gray curves. For comparison with Fig.~\ref{fg:LHC_constraints}, we also depict the favored parameter space for the diphoton excess as gray shaded regions.}
\label{fg:stability_constraints}
\end{figure}
It is time to examine if the electroweak vacuum can be stabilized. To this end, we explore the parameter space for which the electroweak vacuum can be stable up to a high-energy scale. For illustration, we choose several typical energy scales, such as the instability scale $\Lambda^{}_{\rm VS} = 10^{12}~{\rm GeV}$ of the SM vacuum, the GUT scale $\Lambda^{}_{\rm GUT} = 10^{16}~{\rm GeV}$ and the Planck scale $\Lambda^{}_{\rm Pl} = 10^{19}~{\rm GeV}$. The strategy of our numerical analysis is summarized below
\begin{itemize}
\item Following Ref.~\cite{Xing:2011aa}, we first run the SM couplings from $M_Z^{}$ to $m_S^{}$, by using two-loop SM RGEs and adopting one-loop matching for the Higgs quartic coupling $\lambda^{\mathrm{SM}}$ at the top-quark mass threshold. In our calculations, the pole mass of top quark $M^{}_t = 172.9~{\rm GeV}$ is used. As emphasized in Ref.~\cite{Bednyakov:2015sca}, the issue of electroweak vacuum stability is highly sensitive to the top-quark mass, whose experimental uncertainty remains large. This uncertainty should be considered in a more precise study, as the running behavior of $\lambda^{}_H$ will also be changed for a different value of $M^{}_t$ in our case.

\item Then, at $\mu = m_S^{}$, we perform the tree-level matching for the scalar quartic couplings $\lambda_H^{}$, $\lambda_{HS}^{}$ and $\lambda_S^{}$. Such a matching is achieved by solving scalar quartic couplings from Eq.~(\ref{eq:matching_mass}), given the values of $w$ and $\theta$, and the requirements of $m_S = 750~\mathrm{GeV}$ and $m_H^2|_{\mu = m_S^{}} = 2 \lambda^{\mathrm{SM}}|_{\mu = m_S^{}} v^2$.

\item Finally, for a given value of $y_S^{}$, we are able to run all parameters in our model to high energy scales to inspect the vacuum stability problem, with one-loop RGEs derived the previous subsection. It is worthwhile to stress that a dedicated study of the vacuum stabiblity should be carried out with two-loop RGEs and one-loop matching conditions, particularly if the diphoton excess is further confirmed. Learning from the SM case~\cite{Degrassi:2012ry}, we expect that the vacuum stability should be sensitive to high-order radiative corrections. At present, however, the one-loop calculations may be adequate for us to understand how the singlet scalar and vector-like fermions significantly modify the stability of the electroweak vacuum.
\end{itemize}

Requiring the vacuum stability conditions below Eq.~(\ref{eq:condition}) to be satisfied till some high-energy scale and all couplings to stay within the perturbative regions (namely, $\lambda^{}_{H}$, $\lambda^{}_{HS}$, $\lambda^{}_S < 4\pi$, $g_i^{} < \sqrt{4\pi}$ and $y_{t,S} < \sqrt{4\pi}$), we obtain the allowed parameter space for $\sin\theta$ and $y_S^{}$ in Fig.~\ref{fg:stability_constraints}. In consideration of the diphoton production rate, we assume the vector-like fermion $F$ to be a color triplet as before. In Fig.~\ref{fg:stability_constraints}, we take two different values of $w$, i.e., $w = 2$ and $10~\mathrm{TeV}$, and also assume $Q_F^{} = 5/3$ and $1$ as in Fig.~\ref{fg:LHC_constraints}. Three different high-energy scales, at which a deeper electroweak vacuum would be developed, are discussed. The allowed regions for these three cases are indicated by (dark purple) for the Planck scale, (dark purple + purple) for the GUT scale, and lastly (dark purple + purple + pink) for the SM instability scale. We also show the regions that only satisfy the perturbativity requirements for the above three scales, and those favored by the diphoton excess (gray shaded regions). Several observations from Fig.~\ref{fg:stability_constraints} are in order.
\begin{itemize}
\item For $Q_F^{} = 5/3$, we find that the perturbativity and stability requirements till the Planck scale cannot be satisfied at all for the whole parameter space. However, if relaxing such a breaking-down scale to the GUT scale, some allowed parameter space emerges. In contrast, in the case of $Q_F^{} = 1$ we do find the parameter space allowed by both perturbativity and stability till the Planck scale.

\item Comparing with the regions favored by the diphoton excess, one is able to rule out the possibility of $Q_F^{} = 1$ with the perturbtivity requirements alone. However, in the case of $Q_F^{} = 5/3$ a large portion of parameter space favored by the diphoton excess is still retained by perturbativity. Thus, requiring perturbativity alone is impossible to exclude all possibilities of explaining the diphoton excess.

\item Finally, as one can observe from Fig.~\ref{fg:stability_constraints}, the further requirement for vacuum stability sets more stringent limits on the parameter space. In all four cases, the regions of $y_S^{} \lesssim 0.5$ and $\sin\theta \simeq 0.1$ are favored by perturbativity and vacuum stability. This allowed parameter space in fact can be understood by adopting the boundary matching conditions Eqs.~(\ref{eq:matching_mass}) and (\ref{eq:matching_angle}), and the running behaviors of $\lambda$'s. To see this, let us work in the case where $ v/w \ll 1$. From the matching conditions, we then obtain
\begin{eqnarray}
\lambda_H^{} \simeq \frac{m_H^2}{2 v^2} + \frac{\theta^2 m_S^2}{2v^2},\qquad
\lambda_{HS}^{} \simeq \frac{\theta m_S^2}{2 v w}.
\end{eqnarray}
Therefore, if $\theta$ were too small, $\lambda_H$ would receive little tree-level threshold effects, and the positive RG running contribution from $\lambda_{HS}^{}$ is also negligible, so the vacuum stability problem remains as serious as that in the SM. On the other hand, if $\theta$ tends to be too large, say, $\theta \sim 0.3$, this threshold contribution would give $\lambda_H \sim 0.5$ at low energy boundary scale. Such a large boundary value of $\lambda_H^{}$ would easily cause $\lambda_H^{}$ to run towards the Landau pole. As for the favored region of $y_S^{}$, it is mainly due to the fact that a large value of $y_S^{}$ would drive $\lambda_S^{}$ to be negative, exactly in the same way as too large top-quark Yukawa coupling would lead to a negative value of $\lambda^{\mathrm{SM}}$ at high-energy scales.

\item Lastly and most importantly, we observe that even in the scenario of $Q_F^{} = 5/3$ there is no overlap between the regions allowed by both stability and perturbativity and those favored by the diphoton excess. This indicates that within this simple model one is \emph{not} able to simultaneously explain the recently observed diphoton excess and cure the electroweak vacuum stability problem. Therefore, additional degrees of freedom are certainly needed. They could be extra vector-like fermions as discussed in \cite{Chao:2015ttq}, because more vector-like fermions allow for smaller values of $y_S^{}$ to account for the diphoton excess. Alternatively, one may introduce other scalar fields, which may give additional positive contributions to the quartic coupling of the SM Higgs field~\cite{Lebedev:2012zw}. However, the investigations of these possibilities are beyond the scope of current paper.
\end{itemize}
It is also important to point out that the requirement of the stability of the electroweak vacuum could inversely place restrictive bounds on the new physics models that are intended for the diphoton excess~\cite{Mambrini:2015wyu}.

\section{Summary}

The recent observations of diphoton excess from both ATLAS and CMS collaborations have inspired a great number of studies on possible interpretations from new physics beyond the SM. Among them, one simple and straightforward explanation is to identify the diphoton excess as the on-shell production of a singlet scalar $S$ in the $pp \rightarrow S \rightarrow \gamma\gamma$ channel, as well as a pair of vector-like fermions participating in both production and decay processes. Considering the possibility that this singlet scalar field may have non-negligible interactions with the SM Higgs fields, we set out to study the question whether the SM electroweak vacuum instability problem can be cured by this newly introduced scalar field and its associated vector-like fermions. Unfortunately, we find that simultaneously explaining the observed diphoton excess and solving the vacuum stability problem is \emph{not} possible within this simple scenario. Moreover, we have identified the reason behind it, namely, accounting for the diphoton production rates in general requires a large value of the Yukawa coupling between the scalar field and the vector-like fermions, however, such a large value of Yukawa coupling would in turn drive the quartic scalar coupling to be negative, at a scale that is even lower than that present in the SM.

In this respect, among the existing proposals, which explain the diphoton excess but also yield a relatively large value of Yukawa couplings between the introduced scalar fields and fermions, the issue of vacuum instability can be their Achilles' heel. A simultaneous explanation of both diphoton excess and this vacuum instability issue may call for additional complexities from new physics beyond the SM. We are looking forward to deeper thoughts and more experimental data to sort them out.\footnote{Recently, both ATLAS~\cite{MoriondATLAS} and CMS~\cite{MoriondCMS} collaborations have updated their analyses of the diphoton data, and found that the signal at 750 GeV becomes slightly stronger. Although this update does not change our results, it is now more encouraging to study new physics scenarios explaining the diphoton excess and take into account both vacuum stability and perturbativity of relevant coupling constants.}

\section*{Acknowledgements}

This work was supported in part by the Innovation Program of the Institute of High Energy Physics under Grant No. Y4515570U1, by the National Youth Thousand Talents Program, and by the CAS Center for Excellence in Particle Physics (CCEPP).


\newpage

\begin{thebibliography}{99}
\bibitem{ATLAS:2012ae}
  G.~Aad {\it et al.} [ATLAS Collaboration],
  Phys.\ Lett.\ B {\bf 710}, 49 (2012)
  [arXiv:1202.1408].

\bibitem{Chatrchyan:2012tx}
  S.~Chatrchyan {\it et al.} [CMS Collaboration],
  Phys.\ Lett.\ B {\bf 710}, 26 (2012)
  [arXiv:1202.1488].

\bibitem{Aad:2012tfa}
  G.~Aad {\it et al.} [ATLAS Collaboration],
  Phys.\ Lett.\ B {\bf 716}, 1 (2012)
  [arXiv:1207.7214].

\bibitem{Chatrchyan:2012xdj}
  S.~Chatrchyan {\it et al.} [CMS Collaboration],
  Phys.\ Lett.\ B {\bf 716}, 30 (2012)
  [arXiv:1207.7235].

\bibitem{Glashow:1961tr}
  S.~L.~Glashow,
  Nucl.\ Phys.\  {\bf 22}, 579 (1961);
  S.~Weinberg,
  Phys.\ Rev.\ Lett.\  {\bf 19}, 1264 (1967);
  A.~Salam, in {\it Elementary Particle Theory}, edited by N. Svartholm (Almqvist and Wiksells, Stockholm, 1968), p. 367; {\it Weak and Electromagnetic Interactions}, Conf.\ Proc.\ C {\bf 680519}, 367 (1968).

\bibitem{Higgs:1964ia}
  P.~W.~Higgs,
  Phys.\ Lett.\  {\bf 12}, 132 (1964);
  F.~Englert and R.~Brout,
  Phys.\ Rev.\ Lett.\  {\bf 13}, 321 (1964);
  P.~W.~Higgs,
  Phys.\ Rev.\ Lett.\  {\bf 13}, 508 (1964);
  G.~S.~Guralnik, C.~R.~Hagen and T.~W.~B.~Kibble,
  Phys.\ Rev.\ Lett.\  {\bf 13}, 585 (1964);
  P.~W.~Higgs,
  Phys.\ Rev.\  {\bf 145}, 1156 (1966);
  T.~W.~B.~Kibble,
  Phys.\ Rev.\  {\bf 155}, 1554 (1967).

\bibitem{'tHooft:1979bh}
  G.~'t Hooft,
  ``Naturalness, chiral symmetry, and spontaneous chiral symmetry breaking,''
  NATO Sci.\ Ser.\ B {\bf 59}, 135 (1980).

\bibitem{Coleman:1973jx}
  S.~R.~Coleman and E.~J.~Weinberg,
  Phys.\ Rev.\ D {\bf 7}, 1888 (1973).

\bibitem{Nilles:1983ge}
  H.~P.~Nilles,
  Phys.\ Rept.\  {\bf 110}, 1 (1984);
  H.~E.~Haber and G.~L.~Kane,
  Phys.\ Rept.\  {\bf 117}, 75 (1985);
  M.~F.~Sohnius,
  Phys.\ Rept.\  {\bf 128}, 39 (1985).

\bibitem{Cabibbo:1979ay}
  N.~Cabibbo, L.~Maiani, G.~Parisi and R.~Petronzio,
  Nucl.\ Phys.\ B {\bf 158}, 295 (1979);
  P.~Q.~Hung,
  Phys.\ Rev.\ Lett.\  {\bf 42}, 873 (1979);
  M.~Lindner,
  Z.\ Phys.\ C {\bf 31}, 295 (1986);
  M.~Lindner, M.~Sher and H.~W.~Zaglauer,
  Phys.\ Lett.\ B {\bf 228}, 139 (1989);
  P.~B.~Arnold,
  Phys.\ Rev.\ D {\bf 40}, 613 (1989);
  M.~Sher,
  Phys.\ Rept.\  {\bf 179}, 273 (1989);
  Phys.\ Lett.\ B {\bf 317}, 159 (1993)
  [hep-ph/9307342];
  B.~Schrempp and M.~Wimmer,
  Prog.\ Part.\ Nucl.\ Phys.\  {\bf 37}, 1 (1996)
  [hep-ph/9606386].

\bibitem{Altarelli:1994rb}
  G.~Altarelli and G.~Isidori,
  Phys.\ Lett.\ B {\bf 337}, 141 (1994);
  J.~A.~Casas, J.~R.~Espinosa and M.~Quiros,
  Phys.\ Lett.\ B {\bf 342}, 171 (1995)
  [hep-ph/9409458];
  Phys.\ Lett.\ B {\bf 382}, 374 (1996)
  [hep-ph/9603227];
  T.~Hambye and K.~Riesselmann,
  Phys.\ Rev.\ D {\bf 55}, 7255 (1997)
  [hep-ph/9610272];
  G.~Isidori, G.~Ridolfi and A.~Strumia,
  Nucl.\ Phys.\ B {\bf 609}, 387 (2001)
  [hep-ph/0104016];
  G.~Isidori, V.~S.~Rychkov, A.~Strumia and N.~Tetradis,
  Phys.\ Rev.\ D {\bf 77}, 025034 (2008)
  [arXiv:0712.0242];
  J.~Ellis, J.~R.~Espinosa, G.~F.~Giudice, A.~Hoecker and A.~Riotto,
  Phys.\ Lett.\ B {\bf 679}, 369 (2009)
  [arXiv:0906.0954].

\bibitem{Xing:2011aa}
  Z.~Z.~Xing, H.~Zhang and S.~Zhou,
  Phys.\ Rev.\ D {\bf 86}, 013013 (2012)
  [arXiv:1112.3112].

\bibitem{Holthausen:2011aa}
  M.~Holthausen, K.~S.~Lim and M.~Lindner,
  JHEP {\bf 1202}, 037 (2012)
  [arXiv:1112.2415];
  J.~Elias-Miro, J.~R.~Espinosa, G.~F.~Giudice, G.~Isidori, A.~Riotto and A.~Strumia,
  Phys.\ Lett.\ B {\bf 709}, 222 (2012)
  [arXiv:1112.3022];

\bibitem{Degrassi:2012ry}
  G.~Degrassi, S.~Di Vita, J.~Elias-Miro, J.~R.~Espinosa, G.~F.~Giudice, G.~Isidori and A.~Strumia,
  JHEP {\bf 1208}, 098 (2012)
  [arXiv:1205.6497];
  F.~Bezrukov, M.~Y.~Kalmykov, B.~A.~Kniehl and M.~Shaposhnikov,
  JHEP {\bf 1210}, 140 (2012)
  [arXiv:1205.2893];
  S.~Alekhin, A.~Djouadi and S.~Moch,
  Phys.\ Lett.\ B {\bf 716}, 214 (2012)
  [arXiv:1207.0980];
  I.~Masina,
  Phys.\ Rev.\ D {\bf 87}, no. 5, 053001 (2013)
  [arXiv:1209.0393];
  K.~G.~Chetyrkin and M.~F.~Zoller,
  JHEP {\bf 1304}, 091 (2013)
  [arXiv:1303.2890];
  D.~Buttazzo, G.~Degrassi, P.~P.~Giardino, G.~F.~Giudice, F.~Sala, A.~Salvio and A.~Strumia,
  JHEP {\bf 1312}, 089 (2013)
  [arXiv:1307.3536];
  V.~Branchina and E.~Messina,
  Phys.\ Rev.\ Lett.\  {\bf 111}, 241801 (2013)
  [arXiv:1307.5193].

\bibitem{Xing:2014sja}
  Z.~Z.~Xing,
  Int.\ J.\ Mod.\ Phys.\ A {\bf 29}, 1430067 (2014)
  [arXiv:1411.2713].

\bibitem{Zhou:2015kqs}
  Z.~Z.~Xing and S.~Zhou,
  ``Neutrinos in Particle Physics, Astronomy and Cosmology,''
  Springer-Verlag, Berlin Heidelberg (2011);
  S.~Zhou,
  arXiv:1511.07255.

\bibitem{ATLAS1512}
  The ATLAS Collaboration,
  ``Search for resonances decaying to photon pairs in 3.2 fb$^{-1}$ of $pp$ collisions at $\sqrt{s}$ = 13 TeV with the ATLAS detector,''
  ATLAS-CONF-2015-081.

\bibitem{CMS:2015dxe}
  The CMS Collaboration,
``Search for new physics in high mass diphoton events in proton-proton
  collisions at 13TeV,''
  CMS-PAS-EXO-15-004.

\bibitem{Mambrini:2015wyu}
  K.~Harigaya and Y.~Nomura,
  Phys.\ Lett.\ B {\bf 754}, 151 (2016)
  [arXiv:1512.04850];
  Y.~Mambrini, G.~Arcadi and A.~Djouadi,
  arXiv:1512.04913;
  M.~Backovic, A.~Mariotti and D.~Redigolo,
  arXiv:1512.04917;
  A.~Angelescu, A.~Djouadi and G.~Moreau,
  arXiv:1512.04921;
  Y.~Nakai, R.~Sato and K.~Tobioka,
  arXiv:1512.04924;
  S.~Knapen, T.~Melia, M.~Papucci and K.~Zurek,
  arXiv:1512.04928;
  D.~Buttazzo, A.~Greljo and D.~Marzocca,
  arXiv:1512.04929;
  A.~Pilaftsis,
  Phys.\ Rev.\ D {\bf 93}, 015017 (2016)
  [arXiv:1512.04931];
  S.~Di Chiara, L.~Marzola and M.~Raidal,
  arXiv:1512.04939;
  T.~Higaki, K.~S.~Jeong, N.~Kitajima and F.~Takahashi,
  arXiv:1512.05295;
  S.~D.~McDermott, P.~Meade and H.~Ramani,
  arXiv:1512.05326;
  M.~Low, A.~Tesi and L.~T.~Wang,
  arXiv:1512.05328;
  B.~Bellazzini, R.~Franceschini, F.~Sala and J.~Serra,
  arXiv:1512.05330;
  R.~S.~Gupta, S.~J\"{a}ger, Y.~Kats, G.~Perez and E.~Stamou,
  arXiv:1512.05332;
  C.~Petersson and R.~Torre,
  arXiv:1512.05333;
  E.~Molinaro, F.~Sannino and N.~Vignaroli,
  arXiv:1512.05334;
  B.~Dutta, Y.~Gao, T.~Ghosh, I.~Gogoladze and T.~Li,
  arXiv:1512.05439;
  Q.~H.~Cao, Y.~Liu, K.~P.~Xie, B.~Yan and D.~M.~Zhang,
  arXiv:1512.05542;
  S.~Matsuzaki and K.~Yamawaki,
  arXiv:1512.05564;
  A.~Kobakhidze, F.~Wang, L.~Wu, J.~M.~Yang and M.~Zhang,
  arXiv:1512.05585;
  R.~Martinez, F.~Ochoa and C.~F.~Sierra,
  arXiv:1512.05617;
  P.~Cox, A.~D.~Medina, T.~S.~Ray and A.~Spray,
  arXiv:1512.05618;
  D.~Becirevic, E.~Bertuzzo, O.~Sumensari and R.~Z.~Funchal,
  arXiv:1512.05623;
  J.~M.~No, V.~Sanz and J.~Setford,
  arXiv:1512.05700;
  S.~V.~Demidov and D.~S.~Gorbunov,
  arXiv:1512.05723;
  W.~Chao, R.~Huo and J.~H.~Yu,
  arXiv:1512.05738;
  S.~Fichet, G.~von Gersdorff and C.~Royon,
  arXiv:1512.05751;
  D.~Curtin and C.~B.~Verhaaren,
  arXiv:1512.05753;
  L.~Bian, N.~Chen, D.~Liu and J.~Shu,
  arXiv:1512.05759;
  J.~Chakrabortty, A.~Choudhury, P.~Ghosh, S.~Mondal and T.~Srivastava,
  arXiv:1512.05767;
  A.~Ahmed, B.~M.~Dillon, B.~Grzadkowski, J.~F.~Gunion and Y.~Jiang,
  arXiv:1512.05771;
  P.~Agrawal, J.~Fan, B.~Heidenreich, M.~Reece and M.~Strassler,
  arXiv:1512.05775;
  C.~Csaki, J.~Hubisz and J.~Terning,
  arXiv:1512.05776;
  A.~Falkowski, O.~Slone and T.~Volansky,
  arXiv:1512.05777;
  D.~Aloni, K.~Blum, A.~Dery, A.~Efrati and Y.~Nir,
  arXiv:1512.05778;
  Y.~Bai, J.~Berger and R.~Lu,
  arXiv:1512.05779;
  E.~Gabrielli, K.~Kannike, B.~Mele, M.~Raidal, C.~Spethmann and H.~Veerm\"{a}e,
  arXiv:1512.05961;
  R.~Benbrik, C.~H.~Chen and T.~Nomura,
  arXiv:1512.06028;
  J.~S.~Kim, J.~Reuter, K.~Rolbiecki and R.~R.~de Austri,
  arXiv:1512.06083;
  A.~Alves, A.~G.~Dias and K.~Sinha,
  arXiv:1512.06091;
  E.~Megias, O.~Pujolas and M.~Quiros,
  arXiv:1512.06106;
  L.~M.~Carpenter, R.~Colburn and J.~Goodman,
  arXiv:1512.06107;
  J.~Bernon and C.~Smith,
  arXiv:1512.06113;
  W.~Chao,
  arXiv:1512.06297;
  M.~T.~Arun and P.~Saha,
  arXiv:1512.06335;
  C.~Han, H.~M.~Lee, M.~Park and V.~Sanz,
  arXiv:1512.06376;
  S.~Chang,
  arXiv:1512.06426;
  I.~Chakraborty and A.~Kundu,
  arXiv:1512.06508;
  R.~Ding, L.~Huang, T.~Li and B.~Zhu,
  arXiv:1512.06560;
  H.~Han, S.~Wang and S.~Zheng,
  arXiv:1512.06562;
  X.~F.~Han and L.~Wang,
  arXiv:1512.06587;
  M.~X.~Luo, K.~Wang, T.~Xu, L.~Zhang and G.~Zhu,
  arXiv:1512.06670;
  J.~Chang, K.~Cheung and C.~T.~Lu,
  arXiv:1512.06671;
  D.~Bardhan, D.~Bhatia, A.~Chakraborty, U.~Maitra, S.~Raychaudhuri and T.~Samui,
  arXiv:1512.06674;
  T.~F.~Feng, X.~Q.~Li, H.~B.~Zhang and S.~M.~Zhao,
  arXiv:1512.06696;
  O.~Antipin, M.~Mojaza and F.~Sannino,
  arXiv:1512.06708;
  F.~Wang, L.~Wu, J.~M.~Yang and M.~Zhang,
  arXiv:1512.06715;
  J.~Cao, C.~Han, L.~Shang, W.~Su, J.~M.~Yang and Y.~Zhang,
  arXiv:1512.06728;
  F.~P.~Huang, C.~S.~Li, Z.~L.~Liu and Y.~Wang,
  arXiv:1512.06732;
  W.~Liao and H.~Q.~Zheng,
  arXiv:1512.06741;
J.~J.~Heckman,
  arXiv:1512.06773;
 M.~Dhuria and G.~Goswami,
  arXiv:1512.06782;
 X.~J.~Bi, Q.~F.~Xiang, P.~F.~Yin and Z.~H.~Yu,
  arXiv:1512.06787;
 J.~S.~Kim, K.~Rolbiecki and R.~R.~de Austri,
  arXiv:1512.06797;
 L.~Berthier, J.~M.~Cline, W.~Shepherd and M.~Trott,
  arXiv:1512.06799;
  W.~S.~Cho, D.~Kim, K.~Kong, S.~H.~Lim, K.~T.~Matchev, J.~C.~Park and M.~Park,
  arXiv:1512.06824;
  J.~M.~Cline and Z.~Liu,
  arXiv:1512.06827;
  M.~Bauer and M.~Neubert,
  arXiv:1512.06828;
  M.~Chala, M.~Duerr, F.~Kahlhoefer and K.~Schmidt-Hoberg,
  arXiv:1512.06833;
  D.~Barducci, A.~Goudelis, S.~Kulkarni and D.~Sengupta,
  arXiv:1512.06842;
  S.~M.~Boucenna, S.~Morisi and A.~Vicente,
  arXiv:1512.06878;
  C.~W.~Murphy,
  arXiv:1512.06976;
  P.~Athron, D.~Harries, R.~Nevzorov and A.~G.~Williams,
  arXiv:1512.07040;
  A.~E.~C.~Hern¨¢ndez and I.~Nisandzic,
  arXiv:1512.07165;
  U.~K.~Dey, S.~Mohanty and G.~Tomar,
  arXiv:1512.07212;
  G.~M.~Pelaggi, A.~Strumia and E.~Vigiani,
  arXiv:1512.07225;
  J.~de Blas, J.~Santiago and R.~Vega-Morales,
  arXiv:1512.07229;
  A.~Belyaev, G.~Cacciapaglia, H.~Cai, T.~Flacke, A.~Parolini and H.~Ser ôdio,
  arXiv:1512.07242;
  P.~S.~B.~Dev and D.~Teresi,
  arXiv:1512.07243;
  W.~C.~Huang, Y.~L.~S.~Tsai and T.~C.~Yuan,
  arXiv:1512.07268;
  S.~Moretti and K.~Yagyu,
  arXiv:1512.07462;
  K.~M.~Patel and P.~Sharma,
  arXiv:1512.07468;
   M.~Badziak,
  arXiv:1512.07497;
  S.~Chakraborty, A.~Chakraborty and S.~Raychaudhuri,
  arXiv:1512.07527;
  Q.~H.~Cao, S.~L.~Chen and P.~H.~Gu,
  arXiv:1512.07541;
  W.~Altmannshofer, J.~Galloway, S.~Gori, A.~L.~Kagan, A.~Martin and J.~Zupan,
  arXiv:1512.07616;
  M.~Cveti\u{c}, J.~Halverson and P.~Langacker,
  arXiv:1512.07622;
  J.~Gu and Z.~Liu,
  arXiv:1512.07624.


\bibitem{EliasMiro:2012ay}
  J.~Elias-Miro, J.~R.~Espinosa, G.~F.~Giudice, H.~M.~Lee and A.~Strumia,
  JHEP {\bf 1206}, 031 (2012)
  [arXiv:1203.0237].

\bibitem{DiChiara:2014wha}
  S.~Di Chiara, V.~Keus and O.~Lebedev,
  Phys.\ Lett.\ B {\bf 744}, 59 (2015)
  [arXiv:1412.7036].

\bibitem{McDonald:1993ex}
  J.~McDonald,
  Phys.\ Rev.\ D {\bf 50}, 3637 (1994)
  [hep-ph/0702143];
  B.~Patt and F.~Wilczek,
  hep-ph/0605188.

\bibitem{8TeV_gamma_gamma_ATLAS}
  G.~Aad {\it et al.} [ATLAS Collaboration],
  Phys.\ Rev.\ D {\bf 92}, no. 3, 032004 (2015)
  [arXiv:1504.05511].

\bibitem{8TeV_gamma_gamma_CMS}
 CMS Collaboration, ``Search for an Higgs Like resonance in the diphoton mass spectra above 150 GeV with 8 TeV data'', Tech. Rep. CMS-PAS-HIG-14-006;

\bibitem{Franceschini:2015kwy}
  R.~Franceschini {\it et al.},
  arXiv:1512.04933.

\bibitem{Knapen:2015dap}
  S.~Knapen, T.~Melia, M.~Papucci and K.~Zurek,
  arXiv:1512.04928 [hep-ph].

\bibitem{Gunion:1989we}
  J.~F.~Gunion, H.~E.~Haber, G.~L.~Kane and S.~Dawson,
  Front.\ Phys.\  {\bf 80}, 1 (2000).


\bibitem{Heinemeyer:2013tqa}
  S.~Heinemeyer {\it et al.} [LHC Higgs Cross Section Working Group Collaboration],
  arXiv:1307.1347.

\bibitem{Bowen:2007ia}
  M.~Bowen, Y.~Cui and J.~D.~Wells,
  JHEP {\bf 0703}, 036 (2007)
  [hep-ph/0701035].

\bibitem{Bojarski:2015kra}
  F.~Bojarski, G.~Chalons, D.~Lopez-Val and T.~Robens,
  arXiv:1511.08120.

\bibitem{Falkowski:2015iwa}
  A.~Falkowski, C.~Gross and O.~Lebedev,
  JHEP {\bf 1505}, 057 (2015)
  [arXiv:1502.01361].

\bibitem{VLQ_limit}
  G.~Aad {\it et al.} [ATLAS Collaboration],
  Phys.\ Rev.\ D {\bf 91}, no. 11, 112011 (2015)
  [arXiv:1503.05425 [hep-ex]];
   G.~Aad {\it et al.} [ATLAS Collaboration],
  JHEP {\bf 1508}, 105 (2015)
  [arXiv:1505.04306 [hep-ex]];
   V.~Khachatryan {\it et al.} [CMS Collaboration],
  JHEP {\bf 1506}, 080 (2015)
  [arXiv:1503.01952 [hep-ex]];
  V.~Khachatryan {\it et al.} [CMS Collaboration],
  arXiv:1507.07129 [hep-ex].

\bibitem{g1_running}
  M.~Son and A.~Urbano,
  arXiv:1512.08307;
  Y.~Hamada, T.~Noumi, S.~Sun and G.~Shiu,
  arXiv:1512.08984.

\bibitem{Kadastik:2009cu}
  M.~Kadastik, K.~Kannike and M.~Raidal,
  Phys.\ Rev.\ D {\bf 80}, 085020 (2009)
  [arXiv:0907.1894].

\bibitem{Lerner:2009xg}
  R.~N.~Lerner and J.~McDonald,
  Phys.\ Rev.\ D {\bf 80}, 123507 (2009)
  [arXiv:0909.0520].

\bibitem{Gonderinger:2009jp}
  M.~Gonderinger, Y.~Li, H.~Patel and M.~J.~Ramsey-Musolf,
  JHEP {\bf 1001}, 053 (2010)
  [arXiv:0910.3167].

\bibitem{Gonderinger:2012rd}
  M.~Gonderinger, H.~Lim and M.~J.~Ramsey-Musolf,
  Phys.\ Rev.\ D {\bf 86}, 043511 (2012)
  [arXiv:1202.1316].

\bibitem{Xiao:2014kba}
  M.~L.~Xiao and J.~H.~Yu,
  Phys.\ Rev.\ D {\bf 90}, no. 1, 014007 (2014)
  [arXiv:1404.0681].

\bibitem{RGE_general}
  M.~E.~Machacek and M.~T.~Vaughn, Nucl.\ Phys.\ B {\bf 222}, 83 (1983); Nucl.\ Phys.\ B {\bf 236}, 221 (1984); Nucl.\ Phys.\ B {\bf 249}, 70 (1985).

\bibitem{Cao:2016cok}
  Q.~H.~Cao, Y.~Q.~Gong, X.~Wang, B.~Yan and L.~L.~Yang,
  arXiv:1601.06374;
  S.~F.~Ge, H.~J.~He, J.~Ren and Z.~Z.~Xianyu,
  arXiv:1602.01801.


\bibitem{Bednyakov:2015sca}
  A.~V.~Bednyakov, B.~A.~Kniehl, A.~F.~Pikelner and O.~L.~Veretin,
  Phys.\ Rev.\ Lett.\  {\bf 115}, no. 20, 201802 (2015)
  [arXiv:1507.08833].

\bibitem{Chao:2015ttq}
  W.~Chao, R.~Huo and J.~H.~Yu,
  arXiv:1512.05738.


\bibitem{Lebedev:2012zw}
  O.~Lebedev,
  Eur.\ Phys.\ J.\ C {\bf 72}, 2058 (2012)
  [arXiv:1203.0156];
  P.~S.~Bhupal Dev, D.~K.~Ghosh, N.~Okada and I.~Saha,
  JHEP {\bf 1303}, 150 (2013)
  [arXiv:1301.3453];
  E.~J.~Chun, H.~M.~Lee and P.~Sharma,
  JHEP {\bf 1211}, 106 (2012)
  [arXiv:1209.1303];
  C.~S.~Chen and Y.~Tang,
  JHEP {\bf 1204}, 019 (2012)
  [arXiv:1202.5717];
  J.~Jaeckel, M.~Jankowiak and M.~Spannowsky,
  Phys.\ Dark Univ.\  {\bf 2}, 111 (2013)
  [arXiv:1212.3620];
  L.~Basso, O.~Fischer and J.~J.~van Der Bij,
  Phys.\ Lett.\ B {\bf 730}, 326 (2014)
  [arXiv:1309.6086].


\bibitem{MoriondATLAS} M. Delmastro, {\it Diphoton searches in ATLAS}, 51st Rencontres de Moriond EW 2016. See the slides at  https://indico.in2p3.fr/event/12279/session/12/contribution/163/material/slides/1.pdf; The ATLAS Collaboration, ``Search for resonances in diphoton events with the ATLAS detector
at $\sqrt{s} = 13~\text{TeV}$", ATLAS-CONF-2016-018.

\bibitem{MoriondCMS} P. Musella, {\it Search for high mass diphoton resonances at CMS}, 51st Rencontres de Moriond EW 2016, https://indico.in2p3.fr/event/12279/session/12/contribution/218/material/slides/0.pdf; The CMS Collaboration, ``Search for new physics in high mass diphoton events in $3.3~\text{fb}^{-1}$ of proton-proton collisions at $\sqrt{s} = 13~\text{TeV}$ and combined interpretation of searches at 8 TeV and 13 TeV", CMS-PAS-EXO-16-018.

\end{thebibliography}
\end{document}